\documentclass[lettersize, journal, twocolumn]{article}
\usepackage{amsmath,amsfonts, amssymb}
\usepackage{subcaption}
\usepackage{url}
\usepackage{verbatim}
\usepackage{graphicx}
\usepackage{cite}
\usepackage{stfloats}
\usepackage{orcidlink}
\usepackage{hyperref}
\usepackage{xcolor}
\hypersetup{
    colorlinks,
    linkcolor={red!50!black},
    citecolor={blue!50!black},
    urlcolor={blue!80!black}
}
\usepackage{enumitem}
\usepackage{svg}
\usepackage{url}
\usepackage{adjustbox}

\DeclareMathOperator*{\minimize}{\text{minimize}}
\DeclareMathOperator*{\subjto}{\text{s. t.}}

\date{}

\begin{document}

\title{Optimal planning for heterogeneous autonomous teams with precedence and compatibility constraints and its application on power grid inspection with Unmanned Aerial Vehicles}

\renewcommand\footnotemark{}
\renewcommand\footnoterule{}
\author{Antonio Sojo$^{*1}$\,\orcidlink{0000-0003-4469-0990}, Ivan Maza$^1$\,\orcidlink{0000-0003-3502-8372} and Anibal Ollero$^1$\,\orcidlink{0000-0003-2155-2472}
        \footnote{$^1$ The authors are with the GRVC Robotics Lab, University of Seville, Camino de los Descubrimientos, s/n, 41092 Seville, Spain\\
    *asojo@us.es}%
    }

\maketitle

This work has been submitted to the IEEE-TASE for possible publication. Copyright may be transferred without notice, after which this version may no longer be accessible.

\begin{abstract}
In this paper we address the optimal planning of autonomous teams for general purpose tasks including a wide spectrum of situations: from project management of human teams to the coordination of an automated assembly lines, focusing in the automated inspection of power grids.

There exist many methods for task planning. However, the vast majority of such methods are conceived for very specific problems or situations and are often based in certain assumptions and simplifications. Consider for example all the different algorithms developed to solve the Vehicle Routing Problem (VRP) for all the different vehicles and environment characteristics. This means that no robust general planning method exists and that a possible extension of any of them to a more general situation is often not a trivial task.

To address this, we propose a new truly general method ultimately based on a generalization of the Traveling Salesman Problem (TSP). We call this new model the Heterogeneous Multi-worker Task Planning Problem (HMWTPP).
It provides a natural framework to model many situations typical in task planning of all kinds. Task-Worker compatibility, precedence/order and time-windows constraints are already encoded into the HMWTPP while it can be easily extended to include weight capacity or battery per node constraints in an intuitive manner.

Several classical TSP problems included in the TSPLIB library are solved for validation and performance analysis of HMWTPP showing a comparable numerical performance to that of existing models. In addition, a synthetic example modeling an automated assembly line is analyzed to prove the potential capabilities of the HMWTPP in real-life scenarios.

Ultimately, we focus in the computation of the optimal plan of Unmanned Aerial Vehicles (UAVs) specifically in the context of automated inspection of electrical power grids. This adaptation is validated by solving real-life missions for a sample power grid segment in the ATLAS Flight Test Center at Villacarrillo, Spain.

\end{abstract}

\section{Introduction}

In modern times, due to the exponential increase in the number of workers involved, coordination between members of a group or team, human or automata, is key. This is even more important as entire industries are moving towards complete automation due to the advancements in robotics and artificial intelligence. As of today, task planning, at least in the context of Unmanned Aerial Vehicles (UAVs), is either made manually or computed numerically using mathematical models often based on the Traveling Salesman Problem \cite{tspog1832}. The issue with many of these models is that they are either too specific or lack flexibility, making them a not so good choice for problems for which they were not originally designed. Although many of them are have been proved to work efficiently \cite{bektas2004, cheikhrouhou2021, ines2019, sundar2017, jang2017, junfei2020, ribeiro2020} and cover very diverse applications (vehicle routing, transportation and delivery, network connectivity, search and rescue and agriculture to name a few), they are disseminated and no unique model brings all the different aspects of these models into the same umbrella.

In general, many task planning problems can be described using concepts common to all of them. A series of tasks, which can be either physical or abstract, needs to be performed by a single human or automata or team of them in an optimal manner based on a certain metric. In addition, additional constraints can be imposed. Although their classification is difficult, two main categories arise: 
\begin{itemize}
    \item \textbf{Per worker constraints} imposed only to a specific worker. They include weight capacity constraints in the context of package delivery, battery level in the context of UAVs planning or task compatibility constraints. As an example, the same mathematical expressions for this type of constraint can also be used to model worker fatigue levels in a project management context.
    \item \textbf{Global constraints} imposed over the entire team. They include time windows constraints, i.e., constraints over when a task needs to be performed. They also include constraints that forbid specific feasible solutions.
\end{itemize}

As examples, in \cite{Yuan2021} although weight capacity, energy and time windows constraints are fully modeled, only a single worker is considered. In \cite{mathew2015}, although an entire team is considered, it is not energy-aware and the characteristics of the already chosen team members are directly encoded into the model making it impossible to add compatibility constraints. In \cite{alyassi2023}, authors propose a TSP-based model for UAV path planning that it is not only energy-aware but also compatible with autonomous recharging. It is however limited to a single UAV and does not contain precedence/order nor task compatibility constraints.

In this paper, we present a general task optimal planning model based on a generalization of the Traveling Salesman Problem (TSP) called the Multiworker Task Planning Problem (HMWTPP)\footnote{Which is closely related to the Multi-Robot Task Allocation Problem (MRTAP)} that brings many of the aforementioned characteristics into a single mathematical framework. It uses the abstract concept of task and worker to model generic planning problems that are not only useful for robotics but in many generic situations where an autonomous team is required to complete a set of tasks. This model is presented in \autoref{sec:maths}, where all the mathematical foundation is established. We show that an instance of the model can be represented by a directional multigraph in conjunction with a Mixed Integer Linear Problem (MILP) solvable by any off-the-shelf optimization suites. 
We then use it to solve several classical problems from the TSPLIB \cite{tsplib} in \autoref{sec:classical_problems} and the planning of a guitar assembly line in \autoref{sec:guitar_assembly}. This last example includes many of the unique quirks of our model and shows its versatility.

To conclude, we present a more complex problem: the inspection of power grids using UAVs in \autoref{sec:inspection}. We describe the process of the implementation of UAVs in such scenario and select a specific emplacement, the ATLAS Flight Test Center for real-world planning to further test and validate the HMWTPP model. The solution for several instances of problem for the ATLAS power grid are discussed.

\section{The HMWTPP formulation for optimal task planning problems}
\label{sec:maths}

As we will see, the HMWTPP can be described using a weighed directed multigraph (WDM Graph) coupled with a mixed-integer linear problem (MILP). The corresponding MILP is an optimization problem that encodes all the information needed to completely describe the HMWTPP numerically and to compute its solution. Any off-the-shelf optimization solver can be used to find it. 

\subsection{Weighed directed multigraph representation}

Consider a weighted directed multigraph $G = (V, E, W)$ where $V$ are the vertices, $E$ are the edges and $W$ are the weights of $G$, and a team of heterogeneous workers $\mathcal{W}$. 

The set $\mathcal{W}$ contains all the information needed to characterize and mathematically model each of the team workers $w \in \mathcal{W}$. This team is required to complete a list of tasks $\mathcal{T}$ in a cooperative and optimal manner. Each of the tasks $\tau \in \mathcal{T}$ can be performed in $n_\tau$ different approaches ${\mathcal{T}_\tau = \{\tau^{(n)}\}_{n = 1}^{n_\tau}}$, i.e., the different ways a task can be carried out. Each task approach in the set $\hat{\mathcal{T}} = \cup_{\tau \in \mathcal{T}} \mathcal{T}_\tau$ is associated with a vertex $v_\tau^{(n)}\in V$. The set of all the vertices associated with an specific task is $T_\tau \subset V$ and the set containing all of such vertices is $T = \cup_{\tau \in \mathcal{T}} T_\tau\subset V$, called the set of tasked vertices.

To allow the workers to start and finish their route we define the set of vertices $B\subseteq V$ (${B \cap T = \emptyset }$) or the bases. Bases can either represent a physical deployment point or the abstract command to start and finish the route. The number of bases $b \in B$ must be at least one and all workers $w \in \mathcal{W}$ need to have exactly one base $b_w$ assigned to them although one base can be shared by several ($b_{w_1} = b_{w_2}$ for some $w_1, w_2 \in \mathcal{W}$).

Due to the possible heterogeneity of the team, some of the task approaches might be only compatible with a subset of the workers. To easily express this mathematically, we define, for any subset $S$ of $\hat{\mathcal{T}}$ (and $T$ by extension), the $w$-compatible restriction operation $\cdot \,|_w$ that selects the subset of $S$ that is compatible with the worker $w$:
\begin{equation}
    S|_w = \left\{s \in S \, | \,\Gamma_w(s) = 1 \right\},
\end{equation}
where $\Gamma_w$ is the compatibility function defined as:
\begin{equation}
    \Gamma_w(\tau^{(a)}) = \begin{cases}
        1,\quad & \parbox{14em}{if the worker $w$ is compatible with the approach $a$ of task $\tau$.}\\ 
        \\
        0, &\text{ otherwise.}
    \end{cases}
\end{equation}
This allows us to decompose the set of tasks (and tasked vertices) into compatibility subsets:
\begin{equation}
    \hat{\mathcal{T}} = \bigcup_{\forall w \in \mathcal{W}} \hat{\mathcal{T}}\lvert_{w}; \quad T = \bigcup_{\forall w \in \mathcal{W}} T\lvert_{w}
\end{equation}

Notice that in general for any pair of workers $w_1, w_2 \in \mathcal{W}$, ${\hat{\mathcal{T}}\lvert_{w_1} \cap\,\hat{\mathcal{T}}\lvert_{w_2} \ne \emptyset}$ ($T\lvert_{w_1} \cap\,T\lvert_{w_2} \ne \emptyset$) as a task might be compatible with more than one worker.

To represent the transition of the worker $w$ from the task approach $\tau_1^{(a)}$ to the task approach $\tau_2^{(b)}$, we use the edges $E$ of $G$. One edge $e \equiv [v_{\tau_1}^{(a)}, v_{\tau_2}^{(b)}] | w$ is labeled by an ordered pair of vertices $[v_{\tau_1}^{(a)}, v_{\tau_2}^{(b)}]$ and a worker $w$ (which makes up for the multi in multigraph). In a sense, this multigraph can be seen as a layered set of regular graphs, one for each worker, coupled by common vertices.

The set $E$ must encode all the information about task compatibility and possible task transitions. For this reason, not all possible connections between vertices are allowed. Iterating over each of the workers $w$, its  base $b_w$ (remember that some workers may share the same) is connected to each of the vertices of $T|_w$ in both directions. Then, all the vertices from $T|_w$ are connected with each other except for those that are related to the same task, i.e., vertices from each of sets $T_\tau$ cannot be connected with vertices from the same set. This prevents the execution of the same task twice and avoids trivial loops in the graph. Additionally, some tasks may require extra rules for its connections.

An advantage of imposing the compatibility directly into the connections of the graphs versus doing so using external constraints is that the number of edges decreases and therefore the complexity of $G$. As we will see, this directly translates into a simpler problem to solve and a reduced computational time. Considering the case where all tasks can be done in $n_a$ different approaches by all workers, the number of edges in the worst-case scenario is $\propto n_w(n_a n)^2$ (where $n$ is the number of tasks and $n_w$ is the number of workers) which rapidly scales into a numerical nightmare if the problem is too complex and is not solved carefully. Although no proof is given, it is easy to see that this problem is at least NP-Hard as it is a more complex version of the TSP \cite{papadimitriou1977}.

To quantify the costs of completing a task and the transitions between them, each of the edges $e \equiv [u, v]| w \in E$ is provided with a set of weights $\{\Omega^{(\mu)}_e\}_{\mu} \in W$. They include the cost associated with the transition from $u$ to $v$ and the execution of the task at $v$ for the worker $w$.
\begin{equation}
    \Omega^{(\mu)}_{[u, v]| w} = \Delta \omega^{(\mu)}_{[u, v]| w} + \omega^{(\mu)}_{v | w},
\end{equation}
where $\Delta \omega^{(\mu)}_{[u, v]| w}$ is the transition cost and $\omega^{(\mu)}_{v | w}$ the execution cost. The label $\mu$ classifies different types of costs that might be useful for the model, e.g., if a worker is a ground vehicle, then, time and energy costs might be useful cost types. 

The weights $W$ needs to be computed beforehand so their value does not explicitly depend on the worker's path. One way to avoid this limitation is to check whether a certain condition or events happens mid-route and re-plan the routes with such route forbidden as a solution. This is shown in \autoref{subsec:MILP}.

\subsection{Mixed-integer programming problem (MILP) representation}
\label{subsec:MILP}

Although the WDM graph $G$ is fully constructed, we still lack a description of the actual mathematical problem and its solution. Consider one of the workers, how is its route, i.e., the order at which it performs the assigned tasks, described in the context of $G$? It is represented as a cycle $c$ in $G$, an unordered list of unrepeated adjacent edges (which considering the visited vertices, is also a subgraph of $G$) that loops back to its start vertex. Not at all cycles are valid solutions however. Some extra criteria and constraints are still needed. For instance, all valid routes must start and finish at the corresponding base\footnote{Different starting and ending points can be easily defined if one ignores the last edge of an already planed route and manually adds the transition to the end point. This way, internally the route still closes back to the base which simplifies some aspects of the model.}, so all routes that do not fulfill this are automatically discarded.

To numerically represent a cycle $c_w$, we define a set of binary parameters $\mathbf{Z} = \left\{z_{e}\right\}_{e \in E}$ that activates certain edges of $G$ in a route:
\begin{equation}
    z_{[v_1, v_2] | w} = \begin{cases}
    1, \quad & \text{ if } [v_1, v_2]|w \in p_w\\
    0, \quad & \text{ otherwise}
    \end{cases}
\end{equation}
Notice that they are defined for any pair of vertices even if they are not part of an existing edge in $G$. This might be seen as a poor decision, however it helps in the computation of sums over certain subsets of $E$. Any $z$ is fixed at $0$ for all vertices pairs that are not part of an existing edge while the others are left as free parameters that we can set as a solution to our problem.

As already mentioned, to ensure that a given choice of $\mathbf{Z}$ represents a set of valid routes, it needs to fulfill some additional conditions. Let's first introduce some useful notation in a similar fashion to that of \cite{sundar2017}.

For each worker $w$, we define the two directed divergence operators $\nabla_{\pm}^{(w)}$ over any subset of vertices $S \subseteq V$:
\begin{align}
    \nabla^{(w)}_{+} S &=  \sum_{u \in S}\sum_{v \in V / S} z_{[u,v]|w} \\
    \nabla^{(w)}_{-} S &=  \sum_{u \in S}\sum_{v \in V / S} z_{[v,u]|w}
\end{align}
(notice that the double sum might include non-existent edges, however, those terms yield no contribution). These operators measure how many active edges related to the worker $w$ are exiting ($+$) or entering ($-$), i.e, the number of times the worker enters and exits the subset. We also define the absolute divergence operator $\nabla^{(w)} = \nabla^{(w)}_{+} + \nabla^{(w)}_{-} $. By abuse of notation, we can use any of these operators over a single vertex $v\in V$ using the same notation used for sets.

The integral operator $\Sigma^{(w)}$ defined over any subset $S \subseteq V$ as:
\begin{equation}
    \Sigma^{(w)} S = \sum_{v_1, v_2 \in S} z_{[v_1, v_2] | w },
\end{equation}
is also needed. It represents the total number of connections for a worker within $S$.

For a given $u\in V$ and $w \in \mathcal{W}$ we define the following sets:
\begin{equation}
    \delta^{(w)}_{+}(u) = \left\{ v \in V \,| \, [u, v]|w \in E\right\}
\end{equation}
\begin{equation}
    \delta^{(w)}_{-}(u) = \left\{ v \in V \,| \, [v, u]|w \in E\right\}
\end{equation}

With this notation, we can easily formulate the conditions or constraints needed to define valid cycles.

To ensure that all workers $w\in\mathcal{W}$ cycles loop back to their assigned base $b_w\in B \subset V$ we use the bases constraints ($\mathcal{C}_\text{B}$) \autoref{constraint:bases}:
\begin{equation}
    \left\{ \nabla^{(w)}_{+}b_w = \nabla^{(w)}_{-}b_w = y_{w} \right\}_{w \in \mathcal{W}}
    \label{constraint:bases}
\end{equation}
As some of the workers might remain inactive, we introduce new binary parameters ${\mathbf{Y}_B = \left\{y_{w}\in \{0,1\}\right\}_{w \in \mathcal{W}}}$.

Usually, it is simpler to set $y_w = 1$ for $\forall w \in \mathcal{W}$ and simplify all the rest of the problem by eliminating any of the references to $\mathbf{Y}_B$. This however might yield sub-optimal solutions or even create contradictions. Consider the case where there are less tasks than active workers.
  
As all tasks must be completed by a compatible (remember that compatibility is encoded in the edges of the graph) and active worker exactly once in any of the available approaches, we impose the task completion constraints ($\mathcal{C}_\text{T}$) \autoref{constraint:complete} and \autoref{constraint:complete2}:
\begin{equation}
    \left\{\sum_{w \in \mathcal{W}} \nabla^{(w)}_{+}T_\tau = \sum_{w \in \mathcal{W}}\nabla^{(w)}_{-}T_\tau  = 1         \right\}_{\tau \in \mathcal{T}}
    \label{constraint:complete}
\end{equation}
and
\begin{equation}
    \left\{\nabla^{(w)} v = 2 y_{v|w} \right\}_{\substack{w \in \mathcal{W} \\ v \in T|_w}}
    \label{constraint:complete2}
\end{equation}
We also introduce the parameters ${\mathbf{Y}_C = \{y_{v|w}\in \{0,1\}\}_{w \in \mathcal{W};\,v \in T}}$ to select which worker is to complete each of the tasks and define ${\mathbf{Y} = \mathbf{Y}_C \cup \mathbf{Y}_B}$. In a similar fashion to that of $\mathbf{Z}$, they are separated into fixed and free parameters. If a task $\tau$ is incompatible with $w$ ($\Gamma_w(\tau) = 0$), then, $y_{v_\tau|w}$ is fixed at $0$. All the parameters that correspond to compatible associations are left free to the solver. These constraints also enforce continuity between transitions.

Still, it is possible that the routes yielded by a specific selection of $\mathbf{Z}$ and ${\mathbf{Y} = \mathbf{Y}_B \cup \mathbf{Y}_C}$ contains several closed loops that do not share any common vertex. These are called subtours and they are an unwanted side effect that we need to remove from the model. To do so, we need to use some kind subtour elimination constraints ($\mathcal{C}_\text{S}$). In literature, two types are mainly used: the Dantzig–Fulkerson–Johnson \cite{dfj1954} (DFJ SECs) or the Miller-Tucker-Zemlim \cite{mtz1960} subtour elimination constraints (MTZ SECs). Each of them has their own unique pros and cons. 

On the one hand, the DFJ constraints in our model take the following expressions: 
\begin{equation}
    \left\{\Sigma^{(w)} S \le |S|-1 \right\}_{\substack{W \in \mathcal{W} \\ S \in \{Q\subset V \,|\, b_u \notin Q,\,|Q| \ge 2\}}}
    \label{eq:dfj}
\end{equation}
Each of them enforces that for a given subset $Q \subset V$, the number of moves done by worker within the subset is at most the minimum number of moves needed to transverse all of its vertices. This way, no valid route can loop to itself within $Q$. As we know where each of the workers starts and finishes, we do not need to add the condition for all the possible subsets of $V$, just for those that do not contain the base.

The main advantage of these constraints is their simplicity and generality that allow for a more dynamic solving process without any modification to their original constraint's formulation. The number of subtour constraints scale exponentially as so does the number of possible subsets $Q$. For instances where the number of tasked vertices is significant, the solving process with all the necessary constraints may take an absurd amount of computation effort so a different approach is needed. Instead of solving the complete problem, we relax it by removing the DFJ constraints. We then solve this relaxed instance and check for subtours. If subtours are found, we can eliminate them by adding the corresponding DFJ from \autoref{eq:dfj} for all $Q$ that are a subset of the set of vertices within each of the invalid subtours and then iterate again in a branch-and-cut fashion. In many cases, this approach speeds things up by orders of magnitude depending on the specific problem and alleviate some of the computing requirements. However, this heavily depends on the number of iterations one needs to make until the true optimal solution is found. It is also very limited in terms of extensions to the formulation. The DFJ SECs can also be used to forbid specific possible solutions to the problem.

On the other hand, although the process is not straightforward, the MTZ SECs can be easily adapted to this formulation by modifying the originals to the following:
\begin{equation}
    \left\{p_{u|w} - p_{v|w} + 1 \le m_w(1 - z_{[u,v]|w}) \right\}_{\substack{w \in \mathcal{W} \\ u,v \in T\lvert_w}}
    \label{eq:mtz1}
\end{equation}
and by introducing two auxiliary constraints:
\begin{equation}
    \left\{p_{u|w} \le m_w y_{u|w}\right\}_{\substack{w \in \mathcal{W} \\ u \in T\lvert_w}}
    \label{eq:mtz2}
\end{equation}
and
\begin{equation}
    \left\{z_{[b_w, u]|w} \le p_{u|w}\right\}_{\substack{w \in \mathcal{W} \\ u \in \delta^{(w)}_+(b_w)}},
    \label{eq:mtz3}
\end{equation}
where $m_w = |T\lvert_w|$ and $\left\{p_{u|w}\in\mathbb{N}\right\}_{w \in \mathcal{W};\, u \in T\lvert_w} = \mathbf{P}$ are new free parameters that encode the order at which a vertex $u \in T\lvert_w$ is visited within the route of $w \in \mathcal{W}$. The constraints \autoref{eq:mtz1} enforces that if $u$ is visited before $v$ by $w$, then $p_{u|w} \ge p_{v|w}$. Notice however that if $v$ is visited immediately after $v$, $p_{v|w}$ might not be equal to $p_{u|w} + 1$ as the number of vertices visited within a route might be less than $m_w$ so \autoref{eq:mtz2} is trivially fulfilled. At the same time, \autoref{eq:mtz2} and \autoref{eq:mtz3} assigns $p = 0$ for non-visited vertices and at least $p = 1$ to the first one in each of the routes respectively. These constraints can be easily tightened so the values of $\mathbf{P}$ directly match with the visit order by defining $m_w = \sum_{e\in E\lvert_w} z_e$ and linearizing the products of the type $z_e m_w$. This however is not very useful and much of the actual useful characteristic of the MTZ SECs are also available in this formulation.

As the original MTZ SECs, they provide a natural framework for visit order constraints, however, they cannot be used for a relax-solve-iterate solving approach as they all must be included simultaneously to function properly. A mixed approach using both MTZ and DFJ might be feasible however.

Depending on the problem and in the case that the MTZ SECs are used, it might be useful to add constraints into the order of visit of each vertex in a route. For example, consider that the worker $w$ must visit $u$ before $v$ (denoted as $u <_w v$) for some technical reason. To implement this behavior, we would need to impose a new type of constraint in the model, the order of visit constraints $\mathcal{C}_\text{O}$. For each of the workers $w$ and pair of order-constrained vertices $[u, v]$, the following constraint must be added:
\begin{equation}
    p_{u|w} - p_{v|w}  \le m_w(2 - y_{u|w} - y_{v|w})
    \label{eq:ordering}
\end{equation}
It forces that $p_{u|w} \le p_{v|w}$ unless ${y_{u|w} = 0}$ or ${y_{u|w} = 0}$, i.e., either $u$ or $v$ is not visited by $w$ for which case the constraint becomes trivial.

Consider that instead of order constraints imposed directly into specific vertices we need to implement them into task visits in any of the approaches. This is done by imposing \autoref{eq:ordering} in all the pairs of tasked vertices generated by pairing approaches from the first task and from the second one. This can be done for all workers if the limitation is not specific to a worker.

At this point, the model provides no information about the local, per step, status of the workers. This means that although the total cost of a route is known, no information about the partial sum of all the previous visits up to a specific vertex is known. Inspired by \cite{Yuan2021}, some new variables and constraints that keep track of this information are proposed for any monotonically increasing (or decreasing by changing the inequalities) partial cost function $f^{(\mu)}_{w}: V\,\rightarrow \mathbb{R}$ ($\mu$ is any of the cost types, $w\in W$). The following constraints ($\mathcal{C}_\text{MFE}$) are valid and fully determine ${\mathbf{f} = \{f^{(\mu)}_{w}(v) \in \mathbb{R}\}_{w\in W;\, v\in V}}$:
\begin{multline}
    \Big\{f^{(\mu)}_{w}(u) - f^{(\mu)}_{w}(v) + \Omega^{(\mu)}_{[u,v]|w} \\ \le O_w^{(\mu)}(1 - z_{[u,v]|w}) \Big\}_{\substack{w \in \mathcal{W} \\ u,v \in T\lvert_w}}
    \label{eq:mcfc1}
\end{multline}
and the three auxiliary constraint types:
\begin{equation}
    \left\{f^{(\mu)}_{w}(u) \le  O_w^{(\mu)} y_{u|w}\right\}_{\substack{w \in \mathcal{W} \\ u \in T\lvert_w}},
    \label{eq:mcfc2}
\end{equation}
\begin{equation}
    \left\{\Omega^{(\mu)}_{[b_w, u]|w} z_{[b_w, u]|w} \le f^{(\mu)}_{w}(u)\right\}_{\substack{w \in \mathcal{W} \\ u \in \delta^{(w)}_+(b_w)}}
    \label{eq:mcfc3}
\end{equation}
and
\begin{equation}
    \Big\{f^{(\mu)}_{w}(u) \le  \sum_{v_1, v_2 \in V\lvert_w;\, v_2 \ne b_w} \Omega^{(\mu)}_{[v_1, v_2]|w} z_{[v_1, v_2]|w} \Big\}_{\substack{w \in \mathcal{W} \\ u \in T\lvert_w}}
    \label{eq:mcfc4}
\end{equation}
where $O_w^{(\mu)}$ is any prefixed value that fulfills ${O_w^{(\mu)} > \sum_{e \in E\lvert_w} \Omega^{(\mu)}_{e}}$ (it is larger than any possible complete route cost for $w$).
    
Note that these are a generalization of the MTZ SECs \autoref{eq:mtz1}, \autoref{eq:mtz2} and \autoref{eq:mtz3}. This means that they by themselves can be also used as SECs if the correct cost type is selected. They also assign $f^{(\mu)}_{w}(u) = 0$ if $u$ is not visited by $w$.

As an example, consider the time cost type $\mu = t$. As each value of the function $f^{(t)}_{w}(v)$ is encoding the time at which each visited vertex is completed and is therefore monotonically increasing, the previous constraints can be used to obtain its values internally within the solving process and without any external manipulation after the solution is found.

With these values available to the MILP solver, constraints can also be imposed locally as a per step/vertex basis. Consider the aforementioned case where a vertex must be completed before another one is visited. We can now impose this constraint based on timing instead of route order. We denote this condition as $u \preceq v$, i.e. the time at which $u$ is completed is equal or previous to the time at which $v$ is firstly visited, no matter which worker completes any of them (or $u \prec v$ for the strict inequality). This can be imposed using one precedence constraint ($\mathcal{C}_\text{P}$) for each precedence-constrained pair $[u,v]\in V$:
\begin{equation}
    \sum_{w\in W}f^{(t)}_{w}(u) \le  \sum_{w\in W}f^{(t)}_{w}(v) - \omega^{(t)}_{v | w} y_{v | w}
    \label{eq:precedence}
\end{equation}
(although worker-specific precedence constraints can be imposed individually, each of them by itself are equivalent to the order of visit constraints \autoref{eq:ordering} but more computationally costly). Note that as a vertex can only by visited once and that $f = 0$ for any not visited one, only two values of $f$ are relevant in each constraint.

This type of parameters can also be used as formulated to keep track of the partial energy costs (or equivalently, used or remaining battery charge).

With all of this in mind, we define the space of all possible choices of the values for the free parameters ${\mathbf{\Phi} = \mathbf{Z} \cup \mathbf{Y}\cup \mathbf{P}\cup \mathbf{f}}$ that fulfill all of these constraints as the space of feasible solutions $\mathcal{S}_F$. With a description of what a feasible solution is, we need to define a metric that selects the optimal solution out of the entire space $\mathcal{S}_F$. Such metric must be given as a function ${F:\mathcal{S}_F \rightarrow \mathbb{R}}$ that maps each feasible solution $\mathbf{\Phi} \in\mathcal{S}_F$ to a real value that needs to be minimal (for maximization, we just change the sign of the cost function and minimize the resulting function). 

The specific expression for the cost function heavily depends on the application. However, we give an example that covers many cases.

Consider that we require the solution of least time, given by the slowest route. We would have a set of weights $\{\Omega^{(t)}_e\}_{e \in E}$ that encode all the time costs. The cost function will then be:
\begin{equation}
    F(\mathbf{\Phi}) = \max_{w\in\mathcal{W}} \sum_{v_1, v_2 \in V\lvert_w} \Omega^{(t)}_{[v_1, v_2]|w} z_{[v_1, v_2]|w}
    \label{eq:nonlinear}
\end{equation}
This function however is non-linear and cannot be included in MILPs. To deal with this, we define a linearization of the function that preserves its characterization. We introduce a new free real parameter $M_\Sigma$. We then set $M_\Sigma$ as the upper bound for each of the routes' times and minimize its value. Instead of directly minimizing the slowest work time, we so do indirectly via this new parameter. All of this can be done by adding a new set of constraints ($\mathcal{C}_\text{MTM}$):
\begin{equation}
    \Bigg\{ \sum_{v_1, v_2 \in V\lvert_w} \Omega^{(t)}_{[v_1, v_2]|w} z_{[v_1, v_2]|w} \le M_\Sigma \Bigg\}_{w\in\mathcal{W}}
    \label{constraint:minthemax}
\end{equation}
and defining the cost function (Minimize the Maximum or MTM) as:
\begin{equation}
    F_\text{MTM}(\mathbf{\Phi}) = M_\Sigma
    \label{eq:mtm}
\end{equation}
The main advantage of this approach is that it results in a really simple function and it yields the same results as \autoref{eq:nonlinear}.

Once the cost function is defined, the last thing to do is to actually find the optimal solution. This can be done by solving the following MILP:
\begin{multline}
    \minimize_{\mathbf{\Phi}}\, F_\text{MTM}(\mathbf{\Phi})\\ \,\, \subjto \,\,\mathcal{C}_\text{B} \cup \mathcal{C}_\text{T} \cup \mathcal{C}_\text{S} \cup \mathcal{C}_\text{O} \cup \mathcal{C}_\text{MFE} \cup \mathcal{C}_\text{P} \cup\mathcal{C}_\text{MTM}
\end{multline}
This MILP can be solved with any of the off-the-shelf solvers such as CMPLX or SCIP\cite{scip2} or with any of the standard well-established algorithms. We warn however that, as already mentioned, this formulation can be assumed to be at least NP-hard. Some optimizations will be needed in order to solve the problem in a competent time and more so in real-time applications. One possible optimization is the aforementioned dynamical relaxation of the subtour constraints using the DFJ SECs.

\subsection{Example I: Classical TSPs}
\label{sec:classical_problems}

In \autoref{tab:tsplib_solving}, the solving statistics of several TSPLIB problems are shown. The instance files were directly downloaded from \cite{tsplib} and modified to suit our solver format. In general, only one base is used and it is assigned to the first point in the original file. The original distance scaling was maintained and the speed at which workers move is set to $10$ units of distance per second in all cases. All the problems are solved as a regular TSP, with no time cost assigned to the visit process. Then, some new instances are created with new constraints and modifications that are either randomly generated or by hand by grouping vertices by their number within the instance. In some cases, the problem is solved with several workers, with compatibility constraints and with order/precedence constraints. Each instance has been solved $10$ times using both the DJF and the MTZ and then the average solving time for each case has been computed. In the former case, the dynamic relax-solve-iterate solving approach has been used for subroute subsets with cardinality less than $8$ to avoid too long iteration processes. All the variables and constraints related to order and precedence are also removed in those cases. For the MTZ case, the complete model is used. All MILPs have been solved using the default parameters and methods of the SCIP 8.0.4 Optimization Suite. All the computations in this paper were made using custom Python and SCIP code and problem instances available at \cite{asojogithub2024} in a single core of an \textit{AMD Ryzen R}$5$ $5600$\textit{X} at stock clocks and $32$ Gb of DDR4 RAM at $3200$ MHz.

\begin{table*}
\centering
\begin{tabular}{c|cc|ccc|cc|c}
\textbf{instance} &
  \textbf{$n_w$} &
  \textbf{$n_v$} &
  \textbf{$n_\text{cc}$} &
  \textbf{$n_\text{oc}$} &
  \textbf{$n_\text{pc}$} &
  \textbf{$\langle t_\text{DFJ} \rangle$} &
  \textbf{$\langle t_\text{MTZ} \rangle$} &
  \textbf{Cost} \\ \hline\hline
\textit{eil22} & $1$ & $22$ & 0    & 0    & 0    & $250$ ms  & $2.09$ s  & $27.84$ s \\
               &     &      & 0    & $10$ & 0    & -         & $3042$ s  & $33.01$ s \\
               &     &      & 0    & $20$ & 0    & -         & $272$ ms  & $49.47$ s \\
               &     &      & 0    & 0    & $20$ & -         & $8.27$ s  & $49.47$ s \\ \cline{2-9} 
               & $2$ & $22$ & 0    & 0    & 0    & $574$ s   & D.N.F.    & $15.87$ s \\
               &     &      & $21$ & 0    & 0    & $48$ ms   & $757$ ms  & $21.71$ s \\
               &     &      & $21$ & $19$ & 0    & -         & $53$ ms   & $29.28$ s \\
               &     &      & $21$ & $0$  & $10$ & -         & $274$ s   & $21.93$ s \\ \cline{2-9} 
               & $3$ & $22$ & $21$ & 0    & 0    & $6245$ s  & $1142$ s  & $12.34$ s \\ \hline\hline
\textit{eil23} & $1$ & $23$ & 0    & 0    & 0    & $110$ ms  & $8.23$ s  & $47.01$ s \\
               &     &      & 0    & 0    & $21$ & -         & $11.95$ s & $91.86$ s \\ \cline{2-9} 
               & $2$ & $23$ & 0    & 0    & 0    & $884$ s   & D.N.F.    & $27.62$ s \\
               &     &      & $22$ & 0    & 0    & $80$ ms   & $500$ ms  & $39.75$ s \\
               &     &      & $22$ & $20$ & 0    & -         & $48$ ms   & $58.02$ s \\
               &     &      & $22$ & 0    & $11$ & -         & $729.5$ s & $40.63$ s \\ \cline{2-9} 
               & $3$ & $23$ & $22$ & 0    & 0    & D.N.F.    & $5694$ s  & $21.7$ s  \\ \hline\hline
\textit{eil33} & $1$ & $33$ & 0    & 0    & 0    & $1.36$ s  & D.N.F.    & $44.27$ s \\
               &     &      & 0    & 0    & $31$ & -         & $34.30$ s & $61.79$ s \\ \cline{2-9} 
               & $2$ & $33$ & $32$ & 0    & 0    & $850$ ms  & $35.04$ s & $35.51$ s \\
               &     &      & $32$ & $30$ & 0    & -         & $222$ ms  & $52.92$ s \\ \hline\hline
\textit{att48} & $1$ & $48$ & 0    & 0    & 0    & $2.22$ s  & D.N.F.    & $3352$ s  \\ \cline{2-9} 
               & $2$ & $48$ & 47   & 0    & 0    & $3.35$ s  & $101$ s   & $2460$ s  \\
               &     &      & $47$ & $46$ & 0    & -         & $1.03$ s  & $10308$ s \\ \hline\hline
\textit{eil51} & $1$ & $51$ & 0    & 0    & 0    & $10.59$ s & D.N.F.    & $42.89$ s \\ \cline{2-9} 
               & $2$ & $51$ & $50$ & 0    & 0    & $740$ ms  & $68.65$ s & $30.05$ s \\
               &     &      & $50$ & $27$ & 0    & -         & D.N.F.    &      \\
               &     &      & $50$ & $49$ & 0    & -         & $1.34$ s  & $90.35$ s
\end{tabular}
\caption{Solving statistics for several instances of classical TSPLIB problems. $n_w$ denotes the number of workers, $n_v$ the number of vertices, $n_\text{cc}$ the number of compatibility constraints, $n_\text{oc}$ the number of order constraints,$n_\text{pc}$ the number of precedence constraints, $\langle t_\text{DFJ}\rangle$ and $\langle t_\text{MTZ}\rangle$ are the average compute time of 10 runs using either the DFJ or the MTZ SECs. ``Cost" indicates the value of the cost function for the optimal solution. "D.N.F." stands for \textit{Did Not Finish} before the $10000$ s time limit while ``-" indicates that such a case cannot be solved or is not implemented in our code.}
\label{tab:tsplib_solving}
\end{table*}

The cases where just one worker is active and no additional constraint is added reduce to an asymmetric version of the original TSP, i.e., a TSP where two $z$ variables per edge are used. The run times for such cases are one or two order of magnitude slower that the fastest regular TSP exact algorithms \cite{applegate2007} due to the duplicate variables and the lack of optimization.

In general, it can be seen that the use of the DFJ SECs significantly reduces the computation time up to a certain number of vertices. This is due to lack of the $\mathbf{P}$ variables. However, in some cases the use of MTZ is actually better. Consider the case with $n_w = 3$ for the \textit{eil22} and \textit{eil23}. As the number of workers increases, so does the probability of finding subtours in the relaxed problem. Although no proof is given, this can be seen just by doing simple combinatorics on the original TSP version of the problems and checking how the space of feasible solution increases with $n_w$ while keeping track of the fraction of solution with no subtours. As such fraction decreases, so does the chance of reaching a subtour free solution of the relaxed problem in the dynamic DFJ solution process. At some point, the number of relaxed and semi-relaxed instances needed to completely solve the problem is large enough to surpass the time needed to solve the complete problem with MTZ SECs. This is even stronger when considering that the number of possible DFJ constraints increases exponentially while the number of MTZ SECs increases as a polynomial.

It can also be seen that the variable that most affects the complexity of the problem is the number of workers of the instance. Just by going from $1$ to $2$ the compute time jumps several orders of magnitude. In many TSP-based models that consider each worker as a unique entity such as the HMWTPP, the number of variables increases linearly with the number of workers. As the number of potential solutions increases exponentially (as $2^n$ for problems with only binary variables) with the number of variables, the computation time also scales up on average. The addition of another worker might result in several instances not being solved before the $10000$ s time limit.

On the other hand, the compatibility constraints cut the solving time at least by an order of magnitude as it simplifies the problem by eliminating many feasible solutions from original problem. The results for the order and precedence constraints are mixed however. In general, the precedence constraints are harder to solve as the parameters involved are real numbers instead of integer parameters. In both cases however, the problem is solved faster as the number of constrains increases. As it happened with the compatibility constraints, this is due to smaller feasible solution space although instead of having a simpler graph and MILP, it is the pre-processing done by the SCIP solver that eliminates the invalid solutions.

\begin{figure}[h]
        \centering
        \includesvg[width=0.75\linewidth, inkscapelatex=false]{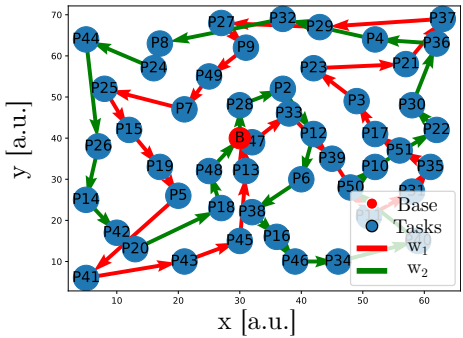}
        \caption{Solution for the \textit{eil51} instance with two workers and 50 compatibility constraints}
        \label{fig:eil51.svg}
\end{figure}

In any case, the main objective of this paper is not the numerical performance nor solution analysis. Several papers already do this in much detail \cite{applegate2007, bektas2004, cheikhrouhou2021, kronqvist2019}. Our point in this section is just to estimate the level of performance of our naive approach to solving the proposed HMWTPP. The computational efficiency obtained in this paper still needs to be improved with state-of-the-art algorithms and a future paper is proposed to address this. It might also be interesting to analyze the performance of non-exact and heuristics algorithms with HMWTPP instances. 

\subsection{Example II: Guitar assembly line with 2 workers}
\label{sec:guitar_assembly}

\begin{figure*}[t]
    \centering
    \begin{subfigure}[b]{0.45\textwidth}
        \centering
        \includesvg[width=0.7\linewidth, inkscapelatex=false]{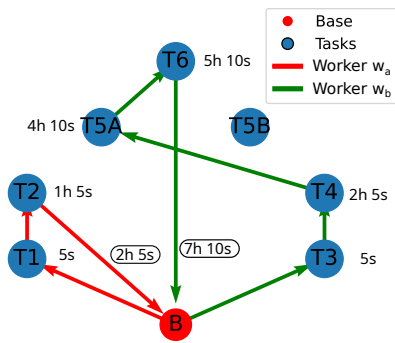}
        \caption{Solution with incomplete formulation. Time: $7$ hours and $10$ seconds.}
        \label{fig:guitar_solution_wrong}
    \end{subfigure}
    \hfill
    \begin{subfigure}[b]{0.45\textwidth}
        \centering
        \includesvg[width=0.7\linewidth, inkscapelatex=false]{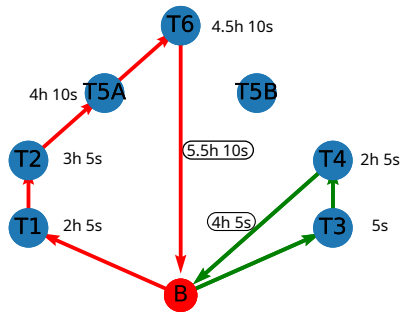}
        \caption{Solution with incomplete formulation. Time: $5$ hours, $30$ minutes and $5$ seconds. Wait time of $1$ hour, $59$ minutes and $55$ seconds}
        \label{fig:guitar_solution_good}
    \end{subfigure}
        \caption{Solutions to the guitar assembly problem in both the incomplete case and the one that includes waiting points.}
        \label{fig:guitar_assembly}
\end{figure*}

To show the versatility of our formulation we will solve a trivial example that models an automated assembly line but includes many of the unique characteristics specific to our model. Consider the manufacturing of a guitar where three task-specific workbenches, WB$1$, WB$2$ and WB$3$ (whose distance is negligible), are used by two different workers, $w_a$ and $w_b$, to complete the steps needed to fully assemble a guitar. The list of tasks, compatibility and time cost for both workers are:
\begin{itemize}
    \item[T1] \label{task:body1} Cut the body at WB1.\\
                {$\Rightarrow$} Time Costs $\rightarrow$ $\{w_a:\,1 \text{ h},\,w_b:\text{ N.C.}\}$
    \item[T2] \label{task:body2} Finish the body for assembly at WB1. \\
                {$\Rightarrow$} Time Costs $\rightarrow$ $\{w_a:\,1 \text{ h},\,w_b:\text{ N.C.}\}$
    \item[T3] \label{task:neck1} Cut the neck at WB2. \\
                {$\Rightarrow$} Time Costs $\rightarrow$ $\{w_b:\text{ N.C.},\,w_a:\,2 \text{ h}\}$
    \item[T4] \label{task:neck2} Finish the neck for assembly at WB2.\\
                {$\Rightarrow$} Time Costs $\rightarrow$ $\{w_b:\text{ N.C.},\,w_a:\,2 \text{ h}\}$
    \item[T5] \label{task:complete} Glue the neck and the body at WB3.\\
                {$\Rightarrow$} Time Costs App. A: $\rightarrow$ $\{w_a:\,0.5 \text{ h},\,w_b:\,1 \text{ h}\}$\\
                {$\Rightarrow$} Time Costs App. B: $\rightarrow$ $\{w_a:\,3 \text{ h},\,w_b:\,3\text{ h}\}$
    \item[T6] \label{task:strings} Finish the guitar at WB3.\\
                {$\Rightarrow$} Time Costs $\rightarrow$ $\{w_a:\,1 \text{ h},\,w_b:\,2 \text{ h}\}$
\end{itemize}
where "N.C." stands for \textit{Not Compatible}. Logically, the following precedence constraints must be imposed: ${\text{T}1 \preceq \text{T}2}$, ${\text{T}3 \preceq \text{T}4}$, ${\text{T}2 \preceq \text{T}5}$, ${\text{T}4 \preceq \text{T}5}$ and ${\text{T}5 \preceq \text{T}6}$. The time taken to move in between different workbenches or the base is always $5$ seconds for both workers. 

The solution given by solving the corresponding MILP using the MTM cost function \autoref{eq:mtm} is represented at \autoref{fig:guitar_solution_wrong} with the slowest route being $7$ hours and $10$ seconds with a computation time of just $50$ ms. This is however not the actual fastest solution.

Due to how the MILP problem is formulated, each worker must leave each task immediately after completion. This means that workers cannot wait for others to complete their tasks. Ideally, the faster worker should wait for the slower at T$2$ and then complete T$5$ and T$6$. Instead, the slowest one is the one who completes the final tasks yielding a slower than optimal solution. 

To fix this, the addition of waiting points in proposed. A waiting point is a new abstract non-mandatory task whose completion time cost $\Omega^\text{WP}_{w} \ge 0$ (one for each worker $w\in W$) is variable and controlled by the MILP. When this or any similar cases are given to the solver, the new variable cost allows the workers to wait a certain amount of time to resume their route, alleviating the aforementioned issue. To keep things simple, these waiting points will be merged into the bases, so no new vertices are added and the only modification needed to be done is to add $\Omega^\text{WP}_{w}$ to all the edges time cost that come out of the base. As the cost function does not explicitly depend on when each task is completed, the waiting process can be done at any point in the route so the value of $\Omega^\text{WP}_{w}$ actually represents the total wait time of the route that that later be distributed between several vertices with enough care. As such, we just need to modify the constrains \autoref{eq:mcfc3} and \autoref{eq:mcfc4} by doing the following transformation: ${f^{(t)}_{w}(v) \rightarrow f^{(t)}_{w}(v) - \Omega^\text{WP}_{w}}$ which translates the time origin for each UAV independently. We also need ensure that the selected value of ${O_w^{(t)}}$ is enough to cover such transformation. When and how much time to wait depends almost exclusively on the user solving the problem with the only restrictions being that the precedence constraints must be fulfilled and the total wait time must be equal to the value given by the solver. 

The solution with this addition is represented at \autoref{fig:guitar_solution_good} and yields a final time of $5.5$ hours and $5$ seconds with a wait time of $1$ hour, $59$ minutes and $55$ seconds, more than $1.5$ hours faster than the previous solution with a computation time of $200$ ms.

The basic HMWTPP is a very flexible and robust framework to model many different planning problems. It is clear however that the specific adaptation to certain problems may require some kind of extension of the original model. We believe however that by using the HMWTPP, this process is much more direct and less complex than the one that would be required by using some other model. With this example, we show that it can be used in a generic assembly line but many other applications exist with none or trivial modifications. Consider for example the scheduling of human resources in a project management setting, the distribution of data within a network, autonomous storage or even autonomous shipment delivery. In the last two cases, the constraints \autoref{eq:mcfc1},  \autoref{eq:mcfc2}, \autoref{eq:mcfc3} and \autoref{eq:mcfc4} would need to be extended to any type of partial cost function, monotonic or not. Then, they can be used to keep track of the physical weight of the packages delivered or moved by the workers and their battery charge. Simultaneously, the HMWTPP already allows the use of time-windows constraints. Consider that we need a package to be delivered to a certain shop within a time windows defined by two fixed time values, otherwise, the delivery is not possible. This is easily done using a constraint of the type:
\begin{equation}
    t_i \le \sum_{w\in W} f^{(t)}_{w}(v) - \omega^{(t)}_{v | w} y_{v | w} \ge t_f
\end{equation}
where $t_i, t_f \in \mathbb{R}$ is the time windows when the delivery in $v$ must be done.

\section{Power grid inspection}
\label{sec:inspection}

One of the main applications of the TSP and its variants is in the vehicle routing problem (VRP) and specifically for UAVs. In this context, several generalization of the TSP and VRP have been used to compute the planning for the inspection of power grids\cite{nekovar2021, ollero2024, caballero2024}. Even though the cited papers show clear success, they lack the flexibility needed to model some of the most crucial aspects of real-life scenarios. Such aspects include the consideration of the physical limitations of fixed-wing aircrafts for the pylon or tower inspection, the possibility of simultaneous planning for power line segments and tower inspection as separate tasks within the same mixed mission or the realistic modeling of UAV autonomy. The last one is critical as they can render planning useless if not correctly done. Current formulations usually use a simplified version of battery behavior where only the total energy consumption per UAV is computed in the planning. This is not an issue for short missions where autonomy is enough to securely power through the complete flight. However, it is not valid for cases where mid-route recharging might be needed. Two approaches are possible to solve this. The first one is to add recharging stations vertices, then solve the problem iteratively, checking whether a solution correctly recharges batteries before depletion and discarding it if not. The second option is more involved but more effective in both potential applications and computationally. The approach used in this work to compute partial costs within the MILP can be extended to non-monotonic functions and used to keep track of the battery level at each route step. Then a constraint on each vertex is added to forbid negative battery levels.

The HMWTPP formulation proposed at \autoref{sec:maths} could be used to solve many of these limitations while keeping a comparable level of numerical performance\footnote{As already said, computational performance is not the focus of this paper and although it is briefly discussed, not much detail is given}. Let's address its implementation in power grid inspection.

\begin{figure*}[t]
     \centering
     \begin{subfigure}[b]{0.45\textwidth}
         \centering
         \includegraphics[width=0.8\textwidth]{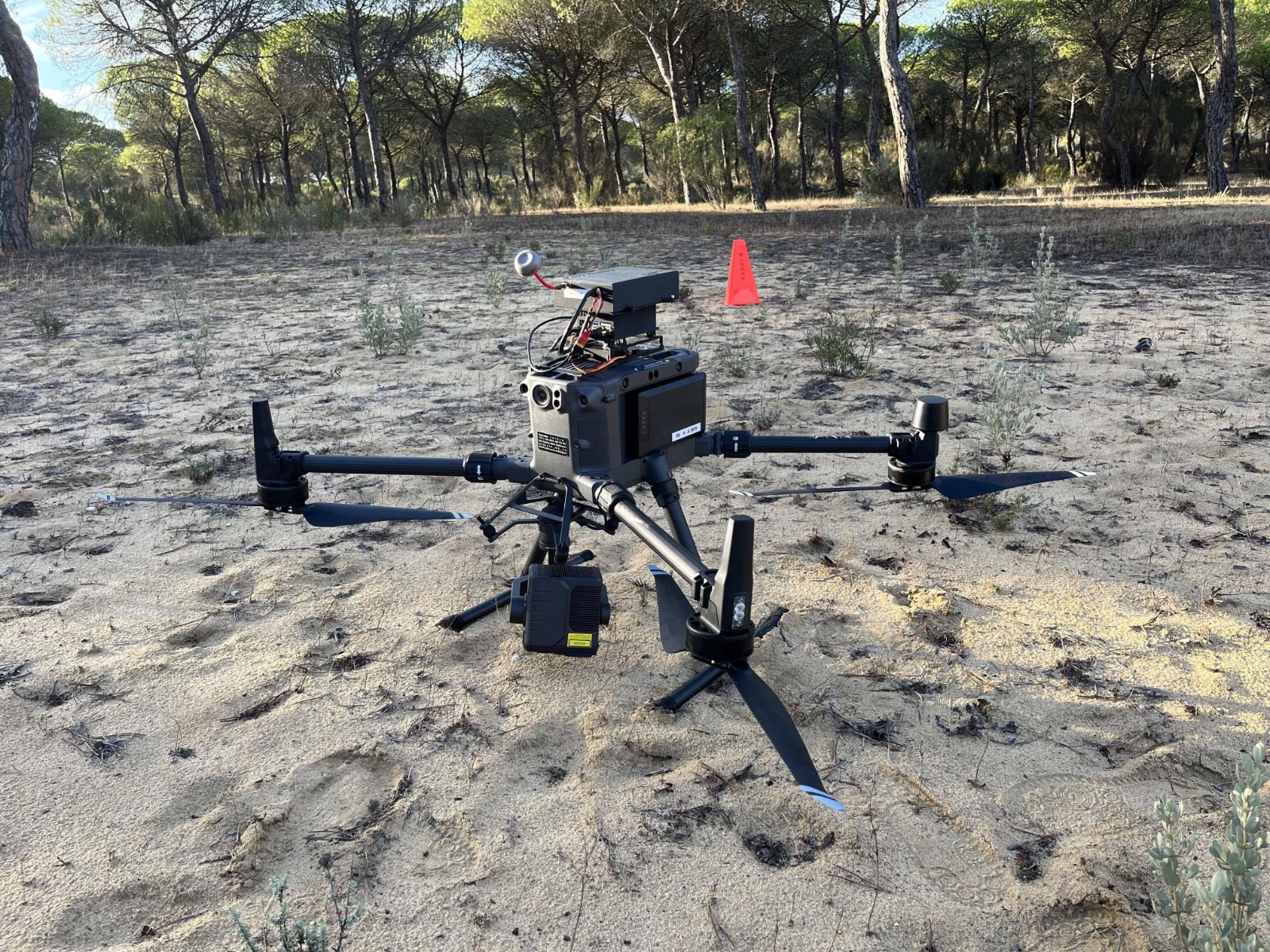}
         \caption{DJI Matrice 300}
         \label{fig:dji_m300}
     \end{subfigure}
     \hfill
     \begin{subfigure}[b]{0.45\textwidth}
         \centering
         \includegraphics[width=0.8\textwidth]{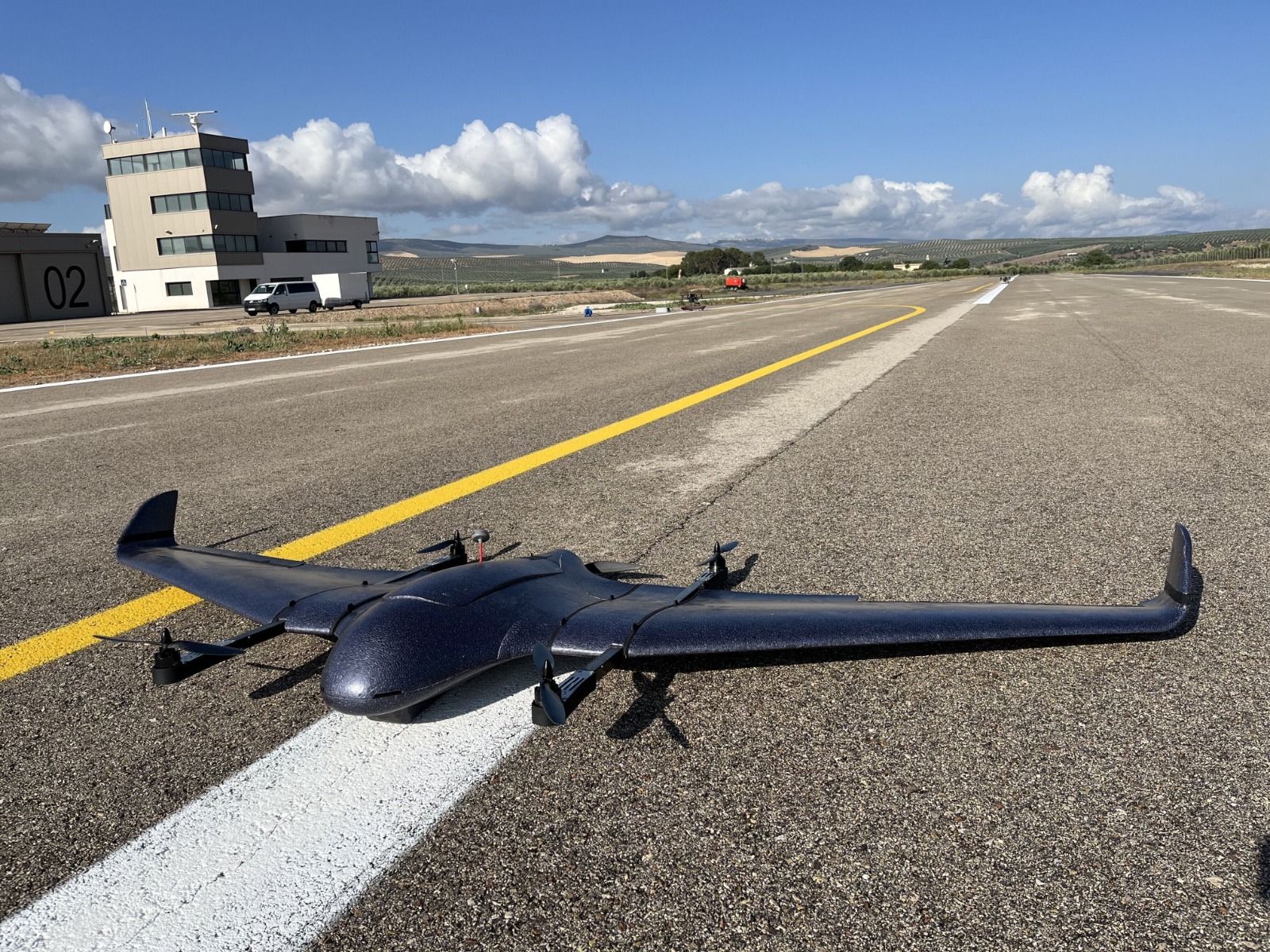}
         \caption{DeltaQuad Pro V2}
         \label{fig:dq_pro_v2}
     \end{subfigure}
        \caption{The two mainly used UAVs for real missions. Photos from \cite{gil-castilla2024}}
        \label{fig:used_uavs}
\end{figure*}

Consider a section of a power grid or an electrical network made of towers (or pylons) and cable segments that connects pairs of towers to each other. We have at our disposal an heterogeneous team of UAVs to inspect a subset of the towers and cable segments as fast as possible. The inspection is done by recording video and taking photos of each of the power grid elements to inspect. The footage is processed either in real-time or after the fact in search of current or potential future failures. 

Each UAV can be represented as a member of the team of workers whose abstract bases are given by their physical landing stations. The autonomy, speeds and physical characteristics and limitations can be easily encoded into the weights and task compatibility of the WDM graph. We mainly consider two types of aircraft: multi-rotor such as the DJI Matrice 300 and fixed-wing (including VTOLs) such as the DeltaQuad Pro V2 VTOL (see \ref{fig:used_uavs}). 

The tasked vertices will correspond to each of the towers and cable segments selected for inspection. UAVs would need to complete a whole circular orbit at a fixed distance around each tower and follow along the length of each cable segment. For each tower, one vertex is added to include the corresponding inspection task. For cable segments, two vertices are needed for the two possible inspection directions. 

The edges follow the general rules already discussed in \autoref{sec:maths}, where the only incompatibility we consider is that VTOL aircrafts cannot inspect towers due to their limited turning radius when in fixed-wing mode.

\begin{figure*}[t]
     \centering
     \begin{subfigure}[b]{0.45\textwidth}
         \centering
         \includegraphics[width=0.8\textwidth]{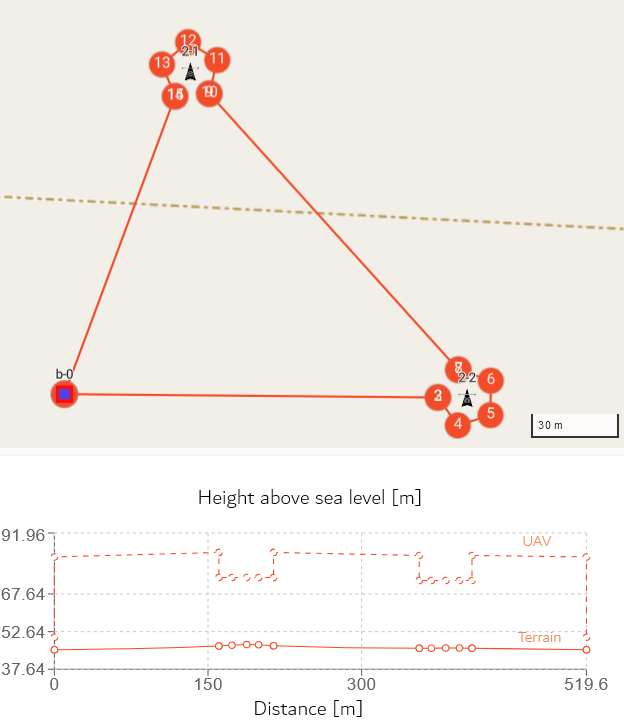}
         \caption{Typical DJI M300 route for tower inspection. $10$ m orbital radius.}
         \label{fig:dji_route}
     \end{subfigure}
     \hfill
     \begin{subfigure}[b]{0.45\textwidth}
         \centering
         \includegraphics[width=\textwidth]{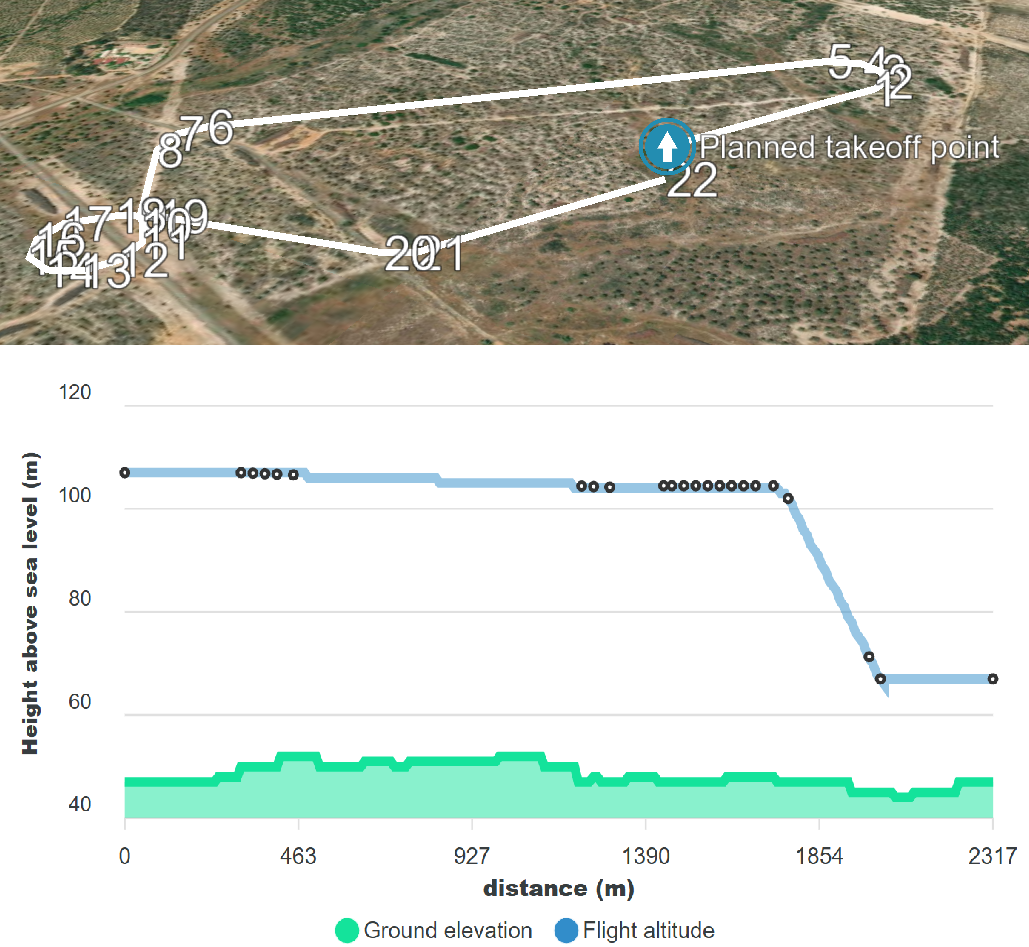}
         \caption{DeltaQuad Pro V2 typical Route using Dubins Paths \cite{validator2024}}
         \label{fig:dq_route}
     \end{subfigure}
        \caption{Typical routes for a DJI M300 (to the left) and a DeltaQuad Pro V2 (to the right)}
        \label{fig:typical_routes}
\end{figure*}

We include two types of weights: $\{\Omega^{(E)}_{e}\}_{e\in E}$ for energy consumption (normalized to unity for each of the UAVs, so $1$ corresponds to their entire energy budget) and $\{\Omega^{(t)}_{e}\}_{e\in E}$ for time costs.\\
For multi-rotors, the energy consumption model considered is the one proposed at \cite{scaramuzza2022} assuming that all the movements between two waypoints are done in a straight line at either navigation or inspection speed. However, in some cases the \textit{eudem25}\cite{eudem25} elevation dataset is used to check for possible terrain collision and to modify the path to avoid it (see \autoref{fig:dji_route}). Usually, this is done by climbing to a safe altitude when moving between inspection tasks and then descending to the inspection altitude once the next waypoint is reached. This, of course adds up to the energy and time costs.\\
For fixed-wing and VTOLs, the same energy model is used while a specific model for such type is being developed by our research group. It is clear that the results obtained after this decision can be nonsensical, however the main points and reasoning remain the same. The trajectories for this type of UAV are computed using the optimal Dubins' paths \cite{dubins1957} between two waypoints. Dubins' paths take into account the limited turning radius to obtain the shortest path between two points with some specific entry and exit directions. These directions in our case are given by the directions of the line segment inspected or by the wind direction in case of takeoff or landing (see \autoref{fig:dq_route}). This wind data is retrieved pre-planning using any of the weather data APIs available, in our case, the OpenWeather API\cite{openweather}. With these types of aircraft, the mission starts with a takeoff againt the wind direction followed by a climb to a safe altitude usually higher than the one for multi-rotors. Possible collisions with terrain are also checked and the path modified if needed.

In addition, the following constraints
\begin{equation}
    \left\{\sum_{u,v\in V\lvert_w} \Omega^{(E)}_{[u,v]|w} z_{[u,v]|w} \le 1\right\}_{w \in W},
\end{equation}
are also added to check whether the UAVs' autonomies are enough for the entire mission execution. They also help with numerical performance as the feasible solution space is considerably reduced.

Consider however the case where no possible route exists to complete the inspection due to autonomy issues. In real applications, two options might be considered to circumvent this. The first option is to segment the inspection into sequences where the team inspects a subset of the structures, then get back to the base, recharge and then continue. However, a more efficient but complex solution exists. A set of charging stations can be distributed throughout the power grid area to allow autonomous recharging. The decision of when to pause the inspection and recharge can be taken either manually or by the solver if we introduce some new constraints to keep track of the battery level at each step of the route in a similar fashion to the MFE constraints $\mathcal{C}_\text{MFE}$ \autoref{eq:mcfc1}, \autoref{eq:mcfc2}, \autoref{eq:mcfc3} and \autoref{eq:mcfc4}\footnote{Notice however that these are not directly valid as battery levels do not have monotonic behavior if charging stations are used}. By adding new non-mandatory tasked vertices to $V$ we can represent the charging stations and process, and by defining the energy costs associated to them as negative, we allow the model to manage autonomous recharging. The amount of charge and time spent at each station can also be a free parameter in the MILP similar to the wait time defined at \autoref{sec:guitar_assembly} or be fixed beforehand. Previous works already do this successfully although in a limited manner \cite{alyassi2023}.

\begin{figure}[h]
         \centering
         \includegraphics[width=0.37\textwidth]{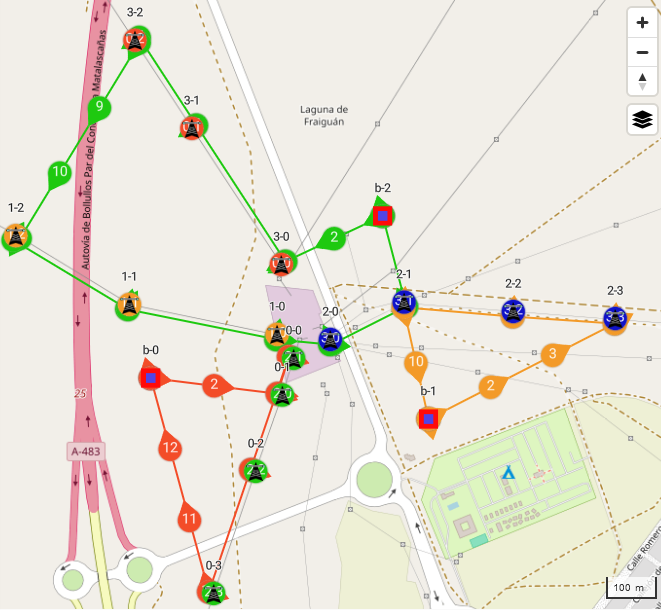}
         \caption{The ground control station displaying the optimal planning for three DJI M300 UAVs inspecting 10 cable segments of a sample power grid near El Rocío, Spain obtained using the HMWTPP model.}
         \label{fig:several_routes}
\end{figure}

Once everything is set up and the MILP is solved, the solution is then parsed into sequences of physical waypoints based on the original paths used to compute the time and energy weights to control the speed, location and the camera gimbal angle (if available) of each UAV. This data is then sent to the ground control station (GCS) and then sent to the corresponding autopilot. In our case, the GCS was developed by \cite{poma2024} (see \autoref{fig:several_routes} for a reference screenshot).

Using the ATLAS Flight Test Center power grid, we define a series of HMWTPP instances\footnote{All the files with the definitions of all instances are available with the code at \cite{asojogithub2024}.}. 4 different UAV models are defined with different navigation and inspection speeds with $2$ different possible bases. For each instance, a random combination of UAV and bases is selected. Wind is generated per instance with random direction and magnitude uniformly with a maximum value of $5$ m/s. For instances with compatibility constraints, one UAV is discarded per task. Following the same procedure from \autoref{sec:classical_problems}, we solve each instance $20$ times, $10$ using DFJ SECs and $10$ using $MTZ$ SECs and then we compute the average solving time. The results are shown in \autoref{tab:atlas_solving}. 
\begin{figure}
    \centering
    \includesvg[width=0.75\linewidth, inkscapelatex=false]{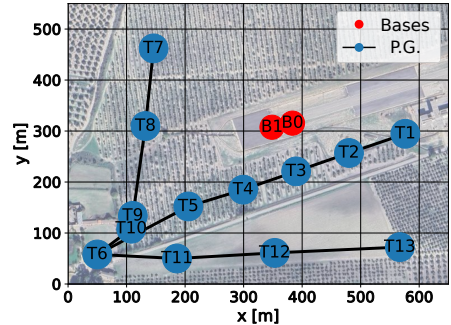}
    \caption{The ATLAS power grid used for testing. The red dot represents the two bases used. The blue dots represent the pylons/towers and the black lines the power cables between them.}
    \label{fig:atlas_pg}
\end{figure}

\begin{table*}
    \centering
\begin{tabular}{c|ccc|c|cc|c}
\textbf{Instance} &
  \textbf{$n_\text{U}$} &
  \textbf{$n_\text{T}$} &
  \textbf{$n_\text{S}$} &
  \textbf{$n_\text{cc}$} &
  \textbf{$\langle t_\text{DFJ} \rangle$} &
  \textbf{$\langle t_\text{MTZ} \rangle$} &
  \textbf{Cost} \\ \hline
  \textit{U1T13} & $1$ & $13$ & 0    & 0    & $18$ ms & $128$ ms & $376.2$ s   \\ \hline
  \textit{EASY} & $2$ & $7$ & 0    & 0    & $370$ ms  & $351$ ms & $173.6$ s   \\
  \textit{U2S12} &   & $0$ & $12$    & 0    & $2.49$ s & $6.68$ s  & $199.4$ s   \\
  \textit{U2T3S3} &   & $3$ & $3$    & 0    & $24.7$ s & $9.71$ s  & $229.2$ s   \\
  \textit{U2T13} &   & $13$ &   0  & 0    & $8.58$ s & $23.37$ s  & $238.7$ s   \\
  \textit{U2T13NC} &   & $13$ &   0  & $13$    & $35$ ms & $300$ ms  & $257.1$ s   \\ \hline
  \textit{U3S12} & $3$  & $0$ &   $13$  & 0    & D.N.F & $192.7$ s  & $147.8$ s   \\
  \textit{U3T3S9} &    & $3$ &   $9$  & 0    & D.N.F & $58.21$ s  & $164.4$ s   \\
  \textit{U3T10S6} &    & $10$ &   $6$  & 0    & D.N.F & $369.54$ s  & $192.01$ s   \\
  \textit{U3T13} &    & $13$ &   $0$  & 0    & D.N.F & $139.3$ s  & $171.6$ s   \\ \hline
  \textit{U4T13} & $4$   & $13$ &   $0$  & $0$    & D.N.F & $133.50$ s  & $107.1$ s   \\
  \textit{U4T13NC} &     & $13$ &   $0$  & $13$    & D.N.F & $53.92$ s  & $117.2$ s \\\hline
  \textit{TOOMANY} & $5$   & $4$ &   $0$  & $0$    & $534$ ms & $645$ ms  & $63.93$ s
       
\end{tabular}
    \caption{Solving statistics for several instances of the ATLAS problem. $n_\text{U}$ denotes the number of UAVs, $n_T$ the number of towers, $n_S$ the number of segments, $n_\text{cc}$ the number of compatibility constraints, $\langle t_\text{DFJ}\rangle$ and $\langle t_\text{DFJ}\rangle$ are the average compute time of 10 runs using the DFJ SECs and MTZ SECs respectively. ``Cost" indicates the value of the cost function for the optimal solution. "D.N.F." stands for \textit{Did Not Finish} before the $10000$ s time limit.}
    \label{tab:atlas_solving}
\end{table*}
The main results are mostly similar to those obtained with the TSPLIB instances in \autoref{sec:classical_problems}. However, some extra nuances appear as we are now using different UAVs in multi-UAV missions and wind is present. In this application, the use of the iterative solving method using DFJ SECs is not that effective. The convergence for $n_\text{U} \ge 3$, i.e., the number of UAVs, is too slow and renders its usage not that effective. It is clear that the use of MTZ SECs is the superior choice in many cases, not only yielding a solution faster but also providing the necessary framework to introduce order and precedence constraints.

Another interesting behavior is the one obtained when solving the same instance with different UAV teams and wind direction and speed. In general, the less symmetric the problem, the faster its solution is found. This can be seen by solving \textit{U4T13} several times, each with a more diverse UAV team and fixed wind speed at $[2,2,0]$ m/s. All the UAVs start at \textbf{B$0$}. The results can be seen in \autoref{tab:u4t13}.

\begin{table*}
    \centering
\begin{tabular}{cc|c|c}
\textbf{Nav. Speeds [m/s]} &
\textbf{Insp. Speeds [m/s]} &
  \textbf{$\langle t_\text{MTZ} \rangle$} &
  \textbf{Cost} \\ \hline
  $[10, 10, 10, 10]$ & $[5, 5, 5, 5]$ &$920$ s & $141.9$ s   \\ 
  $[10, 10, 7, 7]$   & $[5, 5, 5, 5]$ &$678$ s & $159.3$ s  \\
  $[10, 10, 7, 5]$   & $[5, 5, 5, 4]$  & $488.6$ s & $163.6$ s   \\
  $[15, 10, 7, 5]$   & $[15, 5, 5, 4]$  & $79.49$ s & $119.7$ s
       
\end{tabular}
    \caption{Solving statistics for the \textit{U4T13} instance of the ATLAS problem. The results here differ from the original instance as only one base is in use for all UAVs in this computation, in contrast with the first version where both bases are included.}
    \label{tab:u4t13}
\end{table*}

This is similar to what happens when trying to find the global minimum of a multivariate function. If the function is symmetric and flat, the information obtained by moving through space is enough for fast convergence. In this case, many of the branches made by SCIP are very similar and to correctly identify the one with the true solution, everything must be deeply explored.

However, due to the frequent appearance of long branches in the topology of power grids, clustering can be used to improve performance while not losing much information nor detail about the problem. In this context, clustering refers to the association of several physical structures into a single inspection task. As the optimal routes usually follow an entire branch or a big fraction of it from its initial point until the end, a natural clustering rule would be to cluster all the towers and line segments into a single tasked vertex using some kind of heuristics. This would reduce the number of total vertices in the problem thus decreasing the computation time and allowing for easy access to the solution of large-scale problems. Promising results have already been achieved and our own research group has develop a simple yet effective clustering algorithm \cite{roman-escorza2024}.

As a last test, a very simple instance, \textit{TOOMANY}, with $4$ towers and $5$ UAVs is solved. All UAVs are  the same model except for a slower one. In this case, the solver correctly identified and deactivated the slower one, assigning one tower inspection to each of the other UAVs.

This task planner is now part of the autonomous Multi-UAV system developed under the context of the spanish RESISTO Project \cite{resisto} in conjunction with the aforementioned GCS.

\section{Conclusion}
\label{sec:conclusions}

In this paper we address the unification of many of the TSP-based problems used to model the planning of task execution into a single mathematical framework. We explicitly show that our proposed model can be used to compute the optimal route that visits all the points within a graph using several visitors, to compute the optimal plan of autonomous assembly line and to compute the optimal plan for autonomous power grid inspection. It is not difficult to see that it can be also applied in other fields with little to no additional modification. The cover range is wide as it can be used to generalize the genome reconstruction method used in \cite{Agarwala2000} or to compute the Fourier coefficients of the electron density function of a crystal \cite{bland1989} while also providing the ability to compute the optimal route for package delivery with time windows and precedence constraints \cite{malaguti201850} (and with capacity limit if correctly modified). The proposed brings an effective way to introduce TSP-based models into new problems its flexibility allows for easy extensions and modifications.

This flexibility however, comes at a cost: the increase in numerical complexity and poor performance. As the HMWTPP needs to include many different variables to keep track of several per step local parameters, it is a more complex problem than many other TSP-based models. This can be seen in \autoref{sec:classical_problems} where many classical TSPLIB instances have been solved. The time needed to solve these problems is either up to par or several order of magnitude slower than with the use of their TSP and ATSP counterparts and evermore so if heuristic algorithms are used \cite{zhou2018, jiang2020, sarin2014}. This means that optimization is needed and it is proposed as future work. Such future work should analyze the use of heuristics and computation of new inequalities to speed up the HMWTPP solving process. It should also explore its extension to include inequalities for non-monotonic partial cost functions. Another possible extensions for future works are the extension to a mixed Zermelo-TSP \cite{Guerrero2013} approach to allow the model to compute more optimal routes dynamically in very windy environments and the addition of autonomous obstacle avoidance in a UAV Team scenario.

\section{Acknowledgements}
This work was supported by the R+D+i project TED2021-131716B-C22, funded by MCIN/AEI/10.13039/501100011033 and by the European Union NextGenerationEU/PRTR. This work was also supported by Spanish project RESISTO (2021/C005/00144188) funded by FEDER (Fondo Europeo de Desarrollo Regional) from Ministerio de Asuntos Económicos y Transformación Digital.

\bibliographystyle{ieeetr}
\bibliography{bibliography.bib}

\end{document}